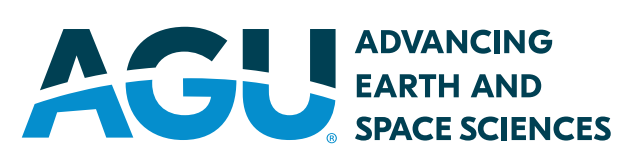

**Key Points:**

- Sabkhas exhibit unique shallow water tables with partial saturation zones occurring directly at the water table level
- Conventional seismic often averages the uppermost complexity, necessitating high-resolution seismic to detect these features
- High-resolution seismic is essential to delineate velocities accurately in complex mature sabkha environments to improve static corrections

**Correspondence to:**
A. Eleslambouly,
100059651@ku.ac.ae;
ak.eleslambouly@gmail.com

**Author Contributions:**
**Conceptualization:** A. Eleslambouly, M. Y. Ali, A. El-Husseiny
**Formal analysis:** A. Eleslambouly, A. El-Husseiny
**Funding acquisition:** M. Y. Ali
**Investigation:** A. Eleslambouly
**Project administration:** M. Y. Ali
**Resources:** A. El-Husseiny, A. A. Al-Shuhail, S. M. Hanafy
**Software:** A. Eleslambouly
**Supervision:** M. Y. Ali, A. El-Husseiny, A. A. Al-Shuhail, F. Bouchalaa
**Validation:** A. Eleslambouly, M. Y. Ali, A. El-Husseiny, A. A. Al-Shuhail, F. Bouchalaa

# Seismic Properties of Coastal and Inland Sabkhas: Implications for Static Corrections

**A. Eleslambouly[1], M. Y. Ali[1], A. El-Husseiny[2], A. A. Al-Shuhail[2], F. Bouchalaa[1], S. M. Hanafy[2], and J. Matsushima[3]**

[1]Department of Earth Sciences, Khalifa University, Abu Dhabi, United Arab Emirates, [2]Department of Geosciences, King Fahd University of Petroleum and Minerals, Dhahran, Saudi Arabia, [3]Graduate School of Frontier Sciences, The University of Tokyo, Tokyo, Japan

**Abstract** Sabkha environments are a prevalent topographic feature in arid coastal areas. Along the Arabian Gulf, sabkhas overlie substantial hydrocarbon reservoirs and exhibit intricate lithological characteristics and an extremely shallow water table. These factors contribute to elevated seismic velocities and signal distortion. Static correction, a crucial initial step in seismic reflection processing, is employed to mitigate the impact of shallow surface layers. In this study, we investigate the variations in seismic properties along the uppermost part of mature and developing sabkhas. We employed high-resolution seismic experiments with geophone spacing of 10 cm to explore the upper tens of centimeters. Conventional surveys with a 2 m spacing complement this approach to investigate deeper layers. Both sabkhas exhibit a unique characteristic of a partially saturated zone, which affects the seismic velocity, leading to lower velocities and consequently influencing the accuracy of the static correction. The high-resolution surveys demonstrated superior accuracy to conventional approaches in determining the top of the partial saturation zone and hardground layer, hence resulting in a more reliable velocity delineation. Moreover, velocities derived from conventional, replacement, and tomogram approaches resulted in unreliable static corrections in mature coastal sabkha compared with developing inland sabkha, attributed to the considerable geological complexity that is characteristic of mature coastal sabkha environments. Carrying out a high-resolution seismic survey in sabkha environments is therefore necessary to mitigate near-surface velocity effects.

**Plain Language Summary** Sabkhas are salt-crusted flat areas found in coastal and inland regions of arid zones, such as the Arabian Gulf. These areas have unique geological characteristics and shallow water tables that pose challenges for seismic surveys to image the subsurface. Traditional seismic methods often overlook important features like partially saturated zones and hardground layers, leading to inaccuracies in the static corrections process associated with variations in the near surface. Our study uses high-resolution and conventional seismic surveys to investigate the seismic properties of both coastal and inland sabkhas. We found that high-resolution seismic is crucial for accurately detecting the velocities of different layers, especially in complex sabkha environments. This method identified a partially saturated zone, which traditional methods often miss, causing errors in static corrections. The presence of evaporitic minerals and hardground layers in mature sabkhas, like those in the mature Abu Dhabi coastal sabkha, leads to higher seismic velocities than developing sabkhas. By comparing velocities obtained from different seismic methods and rock physics models, we confirmed that the observed velocity variations are due to partial saturation rather than lithological changes. Our findings emphasize the importance of using high-resolution seismic surveys in mature sabkha environments to improve the accuracy of seismic studies.

## 1. Introduction

Sabkhas are geomorphic features defined by a flat area with a salt crust overlaying clay, silt, or sand layers (Goodall & Al-Belushi, 1998). They are divided into inland and coastal sabkhas in response to the deflation of sediment surfaces and accumulations in lagoons or a combination of both (Evans, 1970). Coastal sabkhas, prevalent in arid shallow-shelf environments such as the Arabian Gulf, exhibit a direct hydraulic connection with seawater (Evans et al., 1964; Patterson & Kinsman, 1981). In contrast, inland sabkhas are irregularly distributed in creeks, primarily in the northeastern and eastern regions of the Arabian Peninsula, for example, with an indirect relationship to seawater flux (Al-Amoudi et al., 1992; Johnson et al., 1978). Coastal and inland sabkhas show distinct differences in their complexity and composition, each molded by their respective environmental and geological contexts. Coastal sabkhas, directly influenced by marine dynamics, predominantly feature quartz and aragonite as their principal minerals that contribute to creating unique sedimentological features and diverse salt formations. On the other hand, inland sabkhas, which originated in terrestrial sediments, display a varied mineralogical composition, with quartz being the dominant mineral and thin salt beds of gypsum mainly precipitating at and below the water table level (Al-Hurban & Gharib, 2004).

Evaporitic minerals in sabkhas precipitate due to solute availability and intense evaporation, which surpasses precipitation levels (Hussain et al., 2020). Sabkhas are characterized by a shallow water table, typically less than 1 m, facilitating direct evaporation above capillary rise (Yechieli & Wood, 2002). The underlying aquifers in sabkhas are hypersaline; their salinity often exceeds 300 g/L (Alsaaran, 2008; Wood et al., 2002). These brines originate from artesian continental brines and undergo vertical mixing with present-day aquifers (Wood, 2010). Sabkhas serve as outcrop analogs for paleo-sabkhas, which are associated with oil-bearing formations, particularly the Hith and Arab formations in the Arabian Gulf (Alsharhan & Kendall, 1994, 2002). Previous studies on Arabian Peninsula sabkhas employed conventionally spaced geophones, which are insufficient for addressing the heterogeneity in the uppermost geological conditions and accurately mapping the water table, as observed in borehole data (Al-Shuhail & Al-Shaibani, 2009, 2013; El-Hussain et al., 2014).

Sabkhas exhibit pronounced vertical and lateral heterogeneity owing to their complex geological background. A comprehensive understanding of near-surface velocity behavior holds paramount importance for various applications, including static corrections, time-depth conversion, water table depth determination, and geotechnical applications (Al Mesaabi et al., 2012; Baker et al., 1999; Bridle et al., 2007; Hanafy et al., 2020; Soupios et al., 2005). In porous media, compressional velocity ($V_P$) is highly sensitive to presence of gas. Laboratory data show that <2% gas in a brine-saturated sand can lower $V_P$ by more than 40% because the effective bulk modulus of the pore fluid is dominated by gas compressibility (Domenico, 1977; Dvorkin & Nur, 1998). Independently, loose, poorly cemented sands near the surface exhibit intrinsically low frame moduli. Field and laboratory measurements in dunes, beach and aeolian sands typically report $V_P$ in the range 220–1,200 m/s when dry and 1,500–2,000 m/s when fully water-saturated, depending on effective stress (Bachrach et al., 2000; Conte et al., 2009; Hanafy et al., 2020). The interplay of these two mechanisms explains why near surface layers can exhibit velocities of only a few hundred meters per second, whereas fully saturated, more compacted horizons at depth exceed 1,500 m/s (Adelinet et al., 2018; Murad et al., 2014; Pasquet et al., 2015).

Al-Shuhail and Al-Shaibani (2009) used seismic refraction profiling to study the Aziziyah and Ar-Riyas coastal sabkhas of eastern Saudi Arabia, where they mapped the shallow water table because the dry sediments had thicknesses of 0.3 and 1 m, respectively. They found that the average $V_P$ was 300 m/s above the water table in both sabkhas. Below the water table, they found an average $V_P$ of 1,775 m/s in the Aziziyah sabkha and 2,150 m/s in the Ar-Riyas sabkha. Variations in $V_P$ within the top few meters can be quite significant, potentially affecting static corrections (Hanafy et al., 2020). There are no previous studies focused on incorporating high-resolution seismic to resolve the complexity of the uppermost part in sabkhas, which is mostly not captured by conventional seismic surveying. Moreover, to our knowledge, no shallow refraction seismic investigations have been conducted on sabkhas in Abu Dhabi prior to this study. This knowledge gap motivates us to investigate $V_P$ behavior using high-resolution seismic surveys on the coastal sabkha of Abu Dhabi and the sabkha of Jayb Uwayyid along the coastline of the Arabian Gulf. To achieve high-resolution seismic velocities, we implemented a tight (10 cm) geophone spacing, enabling the assessment of velocities in the uppermost tens of centimeters of the sabkhas. These results are complemented by lower-resolution conventional seismic surveys (with a 2 m spacing) to investigate subsurface layers below a depth of 1 m. In doing so, we evaluate the main factors controlling the velocity variations. We used reciprocal, time-term, and tomographic methods to obtain depth $V_P$ profiles and their corresponding implications for static correction accuracy.

## 2. Study Area and General Geology

The first study area, the mature coastal sabkha of Abu Dhabi (CSAD), is situated on the southern shoreline of the Arabian Gulf in the United Arab Emirates (UAE), southwest of Al Rafiq, at the latitude of 24°08′18″N and longitude of 54°06′07″E (Figure 1). This sabkha encompasses various zones influenced by sea-level fluctuations resulting from climatic conditions (Alsharhan & Kendall, 2019). Our seismic experiment focused on the upper supratidal portion, which commenced from the edge of the storm beach ridge and extended into the middle supratidal section. The supratidal zone preserves the most substantial sabkha deposits and a well-developed evaporitic sequence. Regarded as the most extensive coastal sabkha in the Arabian Gulf, the CSAD spans 150 km along the coastline from Ra's Sadr to Tarif, west of Abu Dhabi Island, and reaches 15 km inland (Lokier, 2013). The CSAD represents a complete sabkha, uniquely preserving sequential layers, including lagoon mud (subtidal), microbial mats (subtidal), gypsum mud, and anhydrite nodules (supratidal). This sequence is maintained vertically in the upper supratidal zone, indicative of seaward shallowing, and is underlain by thin hardground primarily composed of skeletal debris cemented by calcite and aragonite (Paul & Lokier, 2017). The sabkha is underlain by a complex sequence of Quaternary sediments that have experienced episodes of deposition and erosion (Kirkham & Evans, 2020).

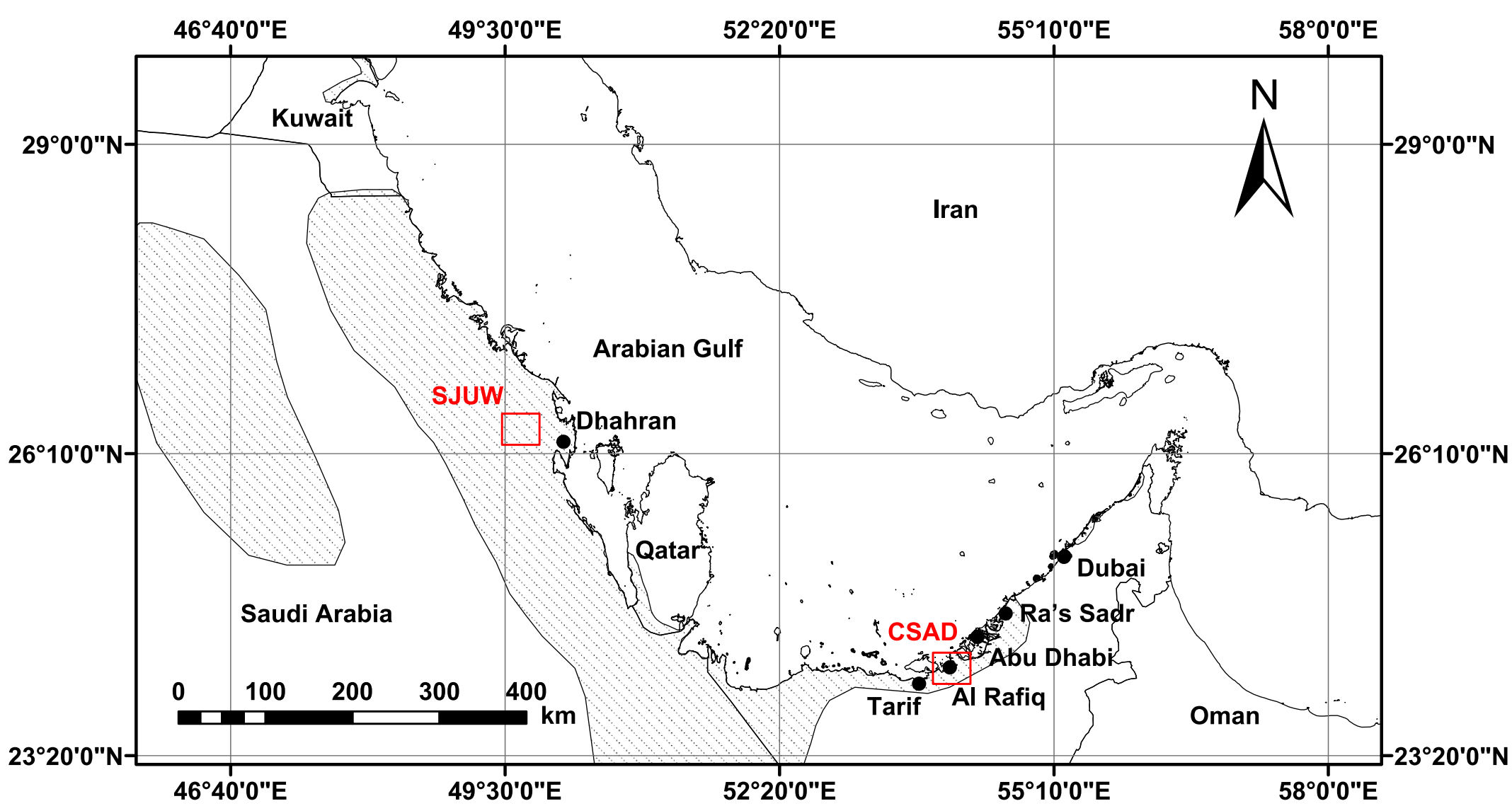

**Figure 1.** Map showing the sabkha distribution (highlighted dashed areas) along the Arabian Gulf (modified after Al-Amoudi et al. (1992)). Red rectangular boxes indicate the two study areas: the coastal sabkha of Abu Dhabi (CSAD) in the UAE and the sabkha Jayb Uwayyid (SJUW) located west of Dhahran city in eastern Saudi Arabia.

The second study area is the developing inland sabkha of SJUW, located in the eastern province of Saudi Arabia, approximately 20 km northwest of Dhahran city (Figure 1). This expansive inland sabkha, one of the largest in the Arabian Peninsula, lies 18 m above sea level and is surrounded by peripheral dunes and sand sheets. The sabkha surface features patches of thin salt crust covered in certain areas by windblown sand. The SJUW exhibits a thin halite crust measuring 3 cm in thickness. The sabkha deposits primarily consist of unconsolidated quartzose sands with poor sorting. Fine- to medium-sized gypsum is present just above the water table (Doebrich & Smith, 1979). SJUW includes a bedded halite deposit located 2.8–4.3 m below the surface, constituting a continuous bed with a thickness ranging from 0.3 to 3.7 m (Dimock, 1955). Uncemented sabkha sediments from the Holocene overlie Pleistocene alternations of sand limestones and shale sequences (Hussain et al., 2020).

## 3. Materials and Methods

### 3.1. Data Acquisition

Seismic refraction surveys covering conventional and high-resolution experiments were conducted at each sabkha, employing distinct acquisition geometries tailored to field logistics and equipment availability (Table 1). During February 2022, a total of 96 channels were deployed at CSAD. The data set was acquired during the peak of shamal winds, where little rainfall occurs, and seasonal flooding occurs, reaching the lower supratidal zone. To increase the spread of geophones, 24 geophones were moved from the starting point and positioned at the end of the survey to cover an additional 46 m, with the survey orientation set in an NW–SE direction perpendicular to the coastline and the primary facies trend (Figures 2a and 2b). A total of 45 shots were performed at CSAD, including two offset shots at each end at 12 and 6 m, respectively. Two high-resolution experiments were also conducted at borehole locations along the seismic profile. Meanwhile, the SJUW seismic survey was conducted in October 2021, and a total of 64 shots were acquired. The geophones were deployed with an E–W orientation extending toward the sand dune slip face (Figure 2c), where the study location had no rain prior to the acquisition date.

**Table 1**
*Acquisition Parameters for the CSAD and SJUW Seismic Surveys*

| Location | CSAD | | SJUW | |
|---|---|---|---|---|
| Survey type | Conventional | High-resolution | Conventional | High-resolution |
| Source | 12 kg sledgehammer | 2.7 kg pointed hammer | 90 kg weight drop | 1.5 kg pointed hammer |
| Channels | 96 | 48 | 72 | 48 |
| Geophones spacing (m) | 2 | 0.1 | 2 | 0.1 |
| Spread length (m) | 190 | 4.7 | 142 | 4.7 |
| Shot interval (m) | 6 | 2.35 | 2 | 2.35 |
| Stacking | 5 | 10 | 3 | 20 |
| Number of Geodes | 4 | 2 | 3 | 2 |
| Total shots | 45 | 3 | 64 | 3 |
| Plate | Metallic 20 × 20 cm | Metallic 10 × 10 cm | Metallic 20 × 20 cm | Metallic 5 × 5 cm |
| Geophones frequency | 8 Hz | | 14 Hz | |
| Sampling (ms) | 0.125 | 0.0625 | 0.5 | 0.0625 |

A single high-resolution survey was acquired at the forward point of the survey using acquisition geometries similar to CSAD but with a smaller weight hammer. The high-resolution surveys had three shooting stations: at the forward (−0.1 m), middle (2.35 m), and reverse points (4.8 m). This approach aims to comprehend the velocity of possible layers that conventional seismic surveys could not have mapped due to resolution differences. Both data sets were collected during a single day visit at each location.

The frequency spectra presented in Figure 3 shows the predominant frequency observed in each survey. These spectra were derived by averaging the frequency spectra of the forward shot gathers. The spectra reveals that the conventional seismic surveys conducted at the CSAD and SJUW are characterized by dominant frequencies of 54 and 58 Hz, respectively. The high-resolution seismic surveys at CSAD display significantly higher dominant frequencies of 209 and 188 Hz. Comparatively, the high-resolution survey at SJUW exhibited an even higher dominant frequency of 281 Hz. This discrepancy in frequencies, particularly the elevated values in the high-resolution survey of SJUW, is primarily attributed to the employment of a smaller weight source, which inherently produces higher frequency signals.

To validate the accuracy of the water table depth estimation through seismic refraction analysis, boreholes were excavated at the midpoint of each high-resolution survey, allowing direct comparison with borehole observations acquired alongside the conventional survey. The recorded lithologies, water table, and hardground level depths at each high-resolution experiment are detailed in Figure 4. Due to the highly cemented nature of the hardground at CSAD, penetration was impossible to the deeper sequences below. Moreover, a sequence analogous to that of the study area in CSAD is observed at the aquaculture trenches (Figure 2a), where the entire stratigraphic sequence is exposed. Samples from the deeper sand layers, as shown in Figure 4, were successfully retrieved from this site. At CSAD, most of the salt facies were concentrated at the upper few centimeters (<70 cm), meanwhile at SJUW, a thin halite bed (4 cm) at the surface and sand with high gypsum was observed below the water table at a depth of 90 cm, and the deeper sequences were almost homogenous sands. In the CSAD borehole, the measured water table is located above the hardground. Overall, CSAD showed high lithological heterogeneity compared to SJUW, where little contrast was observed.

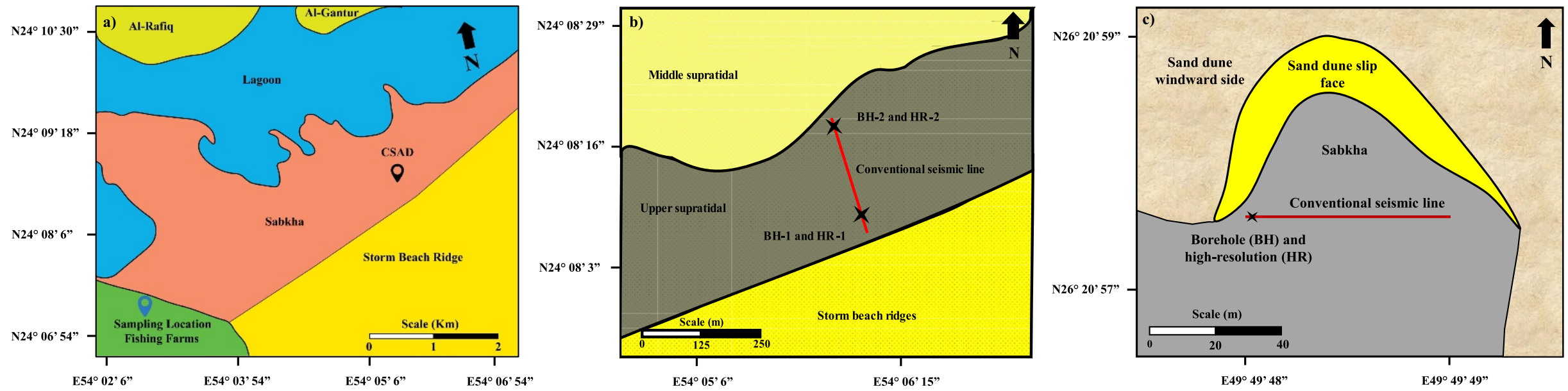

**Figure 2.** Sketches showing the sampling and acquisition sites for the conventional seismic refraction profiles, high-resolution surveys, and locations of drilled boreholes. (a) A map showing locations of the fishing farm trenches (blue marker) and CSAD site (black marker), (b) CSAD survey and BH locations, and (c) SJUW survey and borehole (BH) locations. Refer to Figure 1 for the locations of the CSAD and SJUW sabkhas. BH-1 and BH-2 denote boreholes, and HR-1 and HR-2 represent locations of high-resolution surveys.

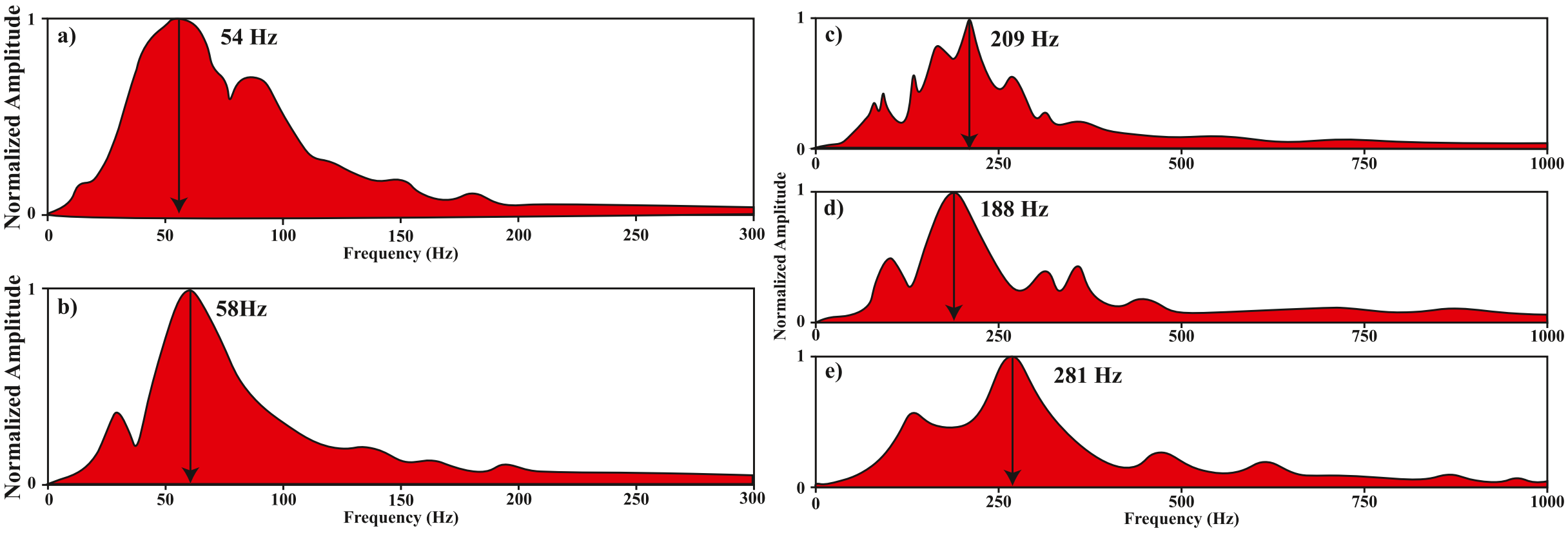

**Figure 3.** The frequency spectrum obtained from head waves arrivals of (a) conventional seismic raw data from CSAD, (b) conventional seismic data from SJUW, (c) high-resolution 1 data from CSAD, (d) high-resolution 2 data from CSAD, and (e) high-resolution survey data from SJUW.

### 3.2. Data Processing

The seismic refraction data, comprising conventional and high-resolution data sets, underwent several processing steps (Figure 5). These steps included automatic gain correction using a window length of 0.1 s to amplify amplitudes at far intervals of the recorded seismograms. Given the low signal-to-noise ratio at far-offset distances, a selective bandpass filter ranging window from 10 to 20 Hz for high pass and 80–120 Hz for low pass was employed on the conventional seismic data to enhance the accuracy of first arrival picking at far noisy geophones. Trace editing included muting noisy channels (poorly planted or contaminated by power line noise) and inverting the polarity of the geophone. Linear interpolation was applied in cases of dead traces.

The first arrivals were manually picked from both conventional (Figure 6) and high-resolution (Figure 7) surveys to construct 2D $V_P$ models. The first arrivals of the CSAD conventional survey (Figure 6a) were subject to the shingling effect, causing a decrease in the amplitude till it fades out; thus, we picked the second head wave arrivals because they are more continuous and representative of the deeper, thicker layer (Du et al., 2021). Sun and Zhang (2013) describe a similar phenomenon of shingled arrivals in seismic refraction data, where sequential, parallel branches of wave arrivals with consistent time shifts are observed. Unlike typical shot records, where the first arrivals display an increasing linear moveout with offset, shingled arrivals show one or more branches that gradually fade into longer offsets, making it difficult to track the first arrivals. In a shot gather, these first arrivals often take on a pattern resembling a Christmas tree, further complicating interpretation. Sun and Zhang (2013) interpret this pattern as evidence of shallow refractions from a thin, high-velocity layer. By examining the slope changes of the arrival time–distance curves (Figures 8 and 9), the main layers constituting the investigated zone were identified based on best fitting to the regression curves.

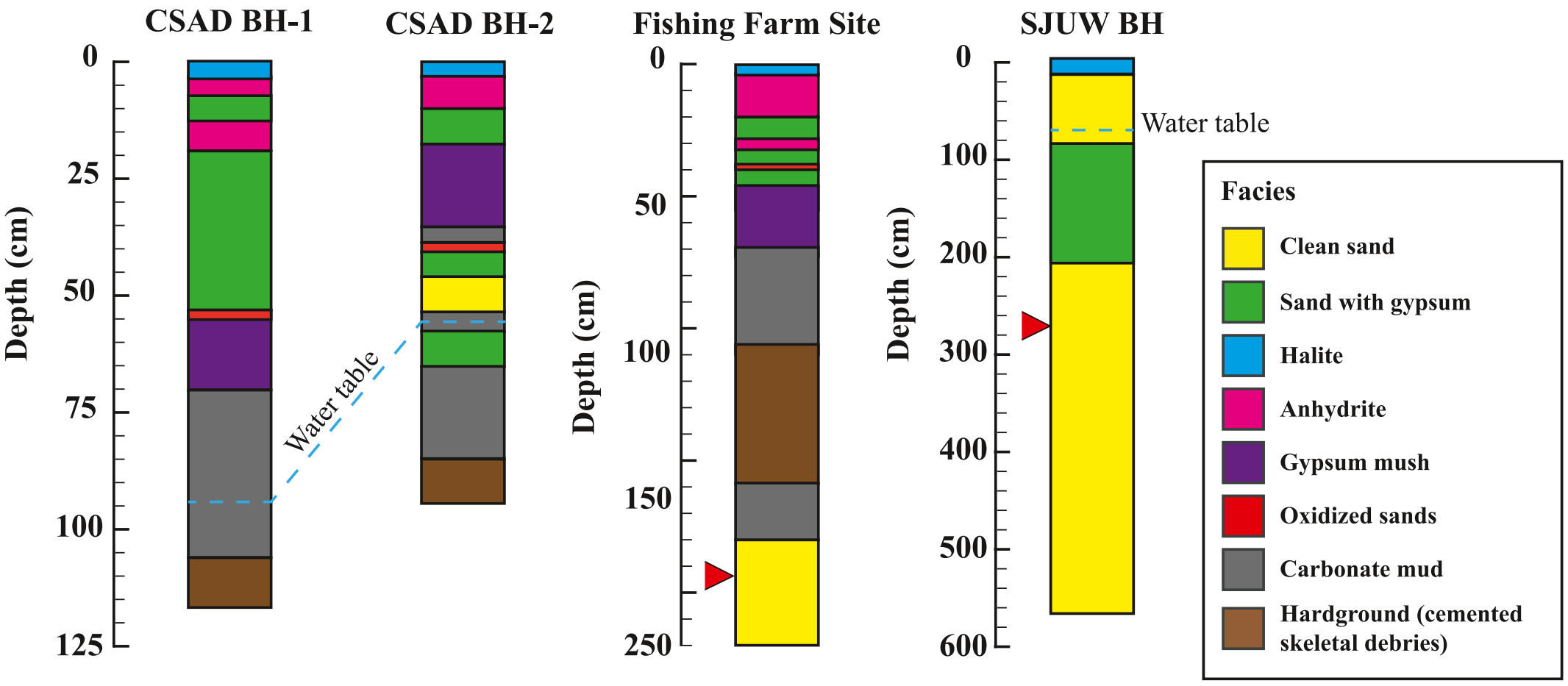

**Figure 4.** Stratigraphic columns of the shallow boreholes and trench at the fishing farm site showing the recorded facies and levels of the water table (dashed blue lines). Refer to Figure 2 for borehole and fishing farm locations. Red triangles indicate the obtained sample depth for rock physics analysis.

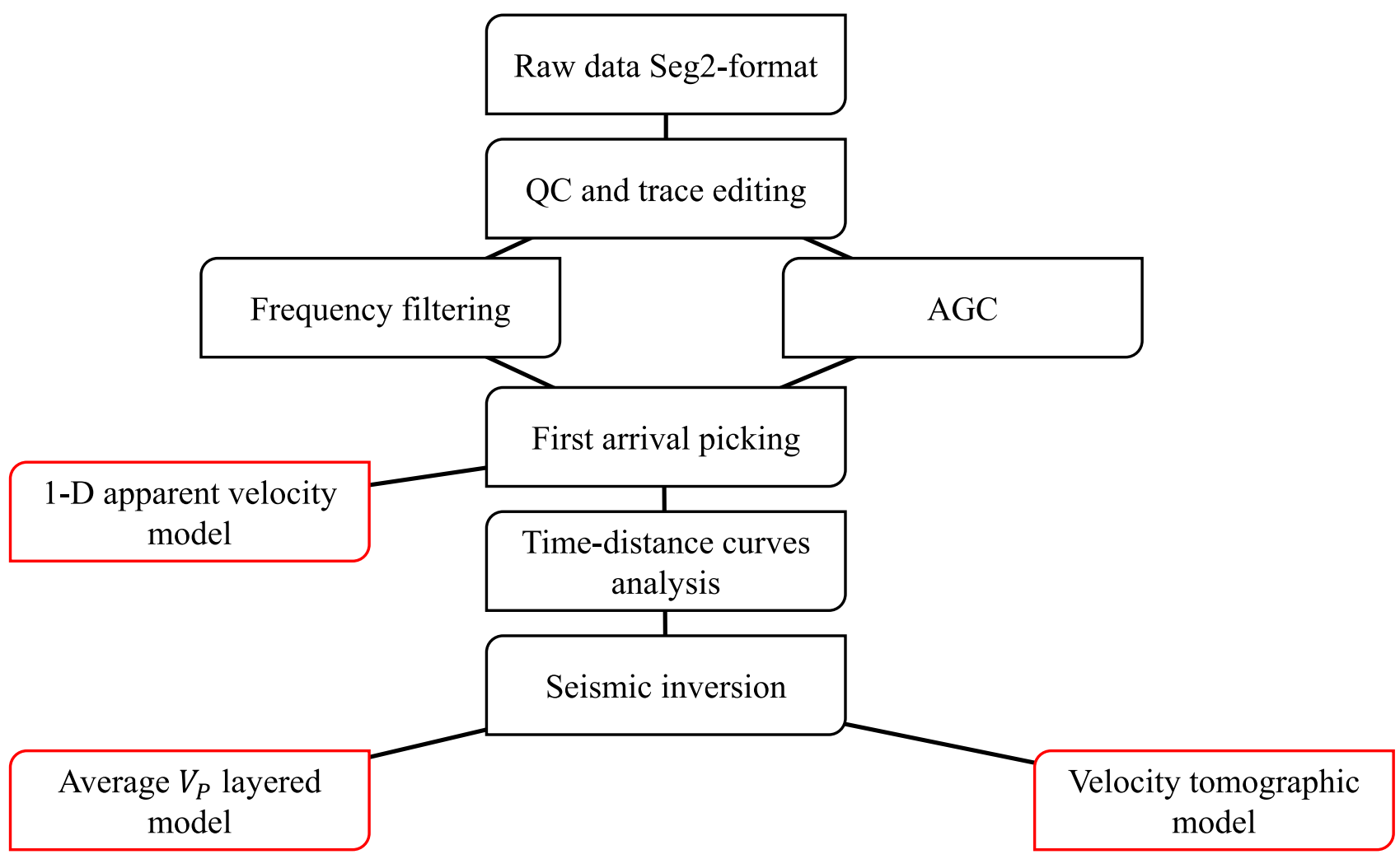

**Figure 5.** Seismic processing and analysis workflow highlighting exported models at various stages (red boxes). Frequency filtering and AGC are not applied at high-resolution experiments scale.

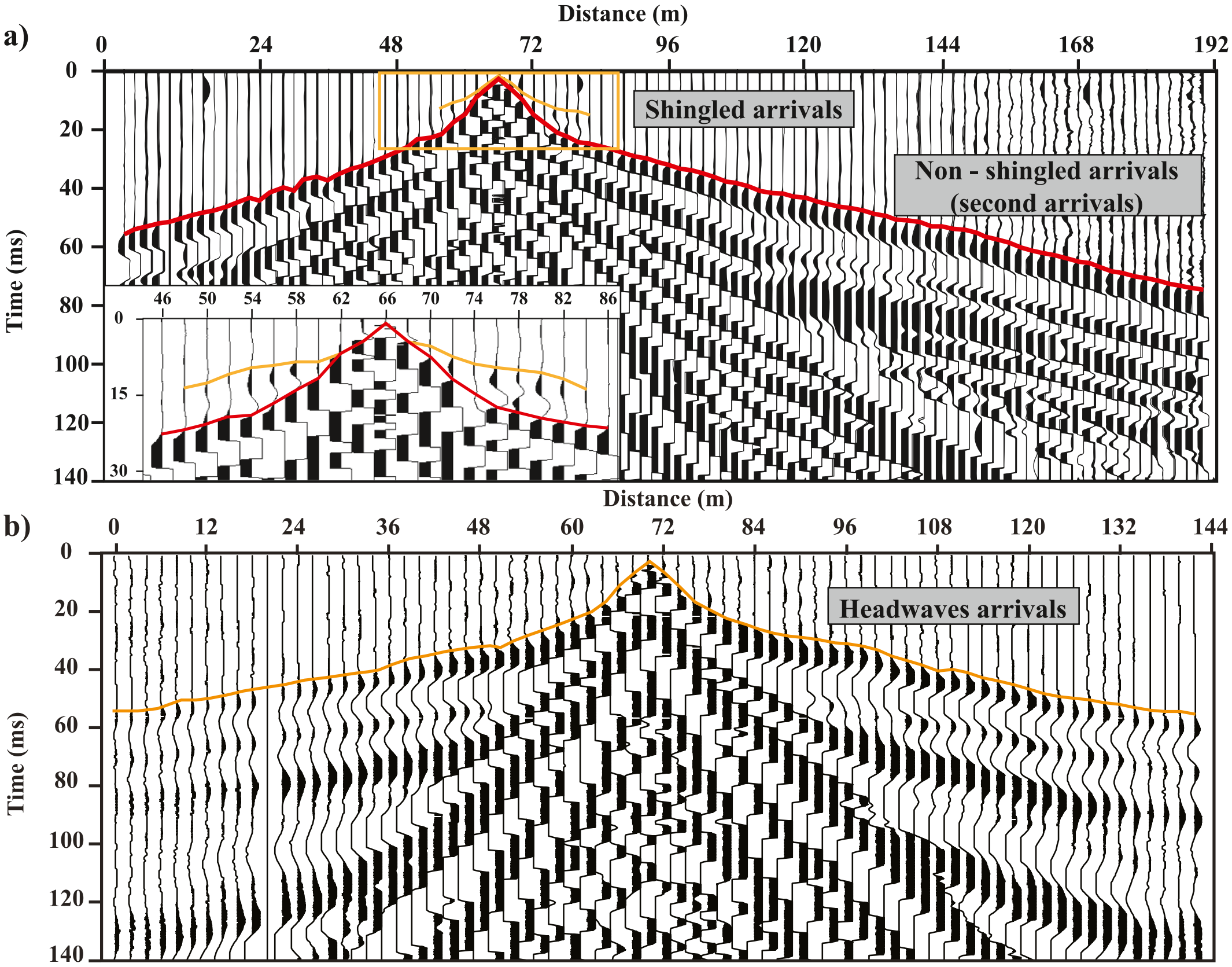

**Figure 6.** Processed shot gathers. (a) At an offset of 66 m from the CSAD conventional seismic survey. The red lines highlight first arrivals, while the orange lines indicate head waves (shingled), discontinuous arrivals in CSAD, and continuous arrivals in SJUW cases. The bottom left corner presents a magnified cropped view of the shingled arrivals. (b) Mid-point shot gather from the SJUW conventional seismic survey.

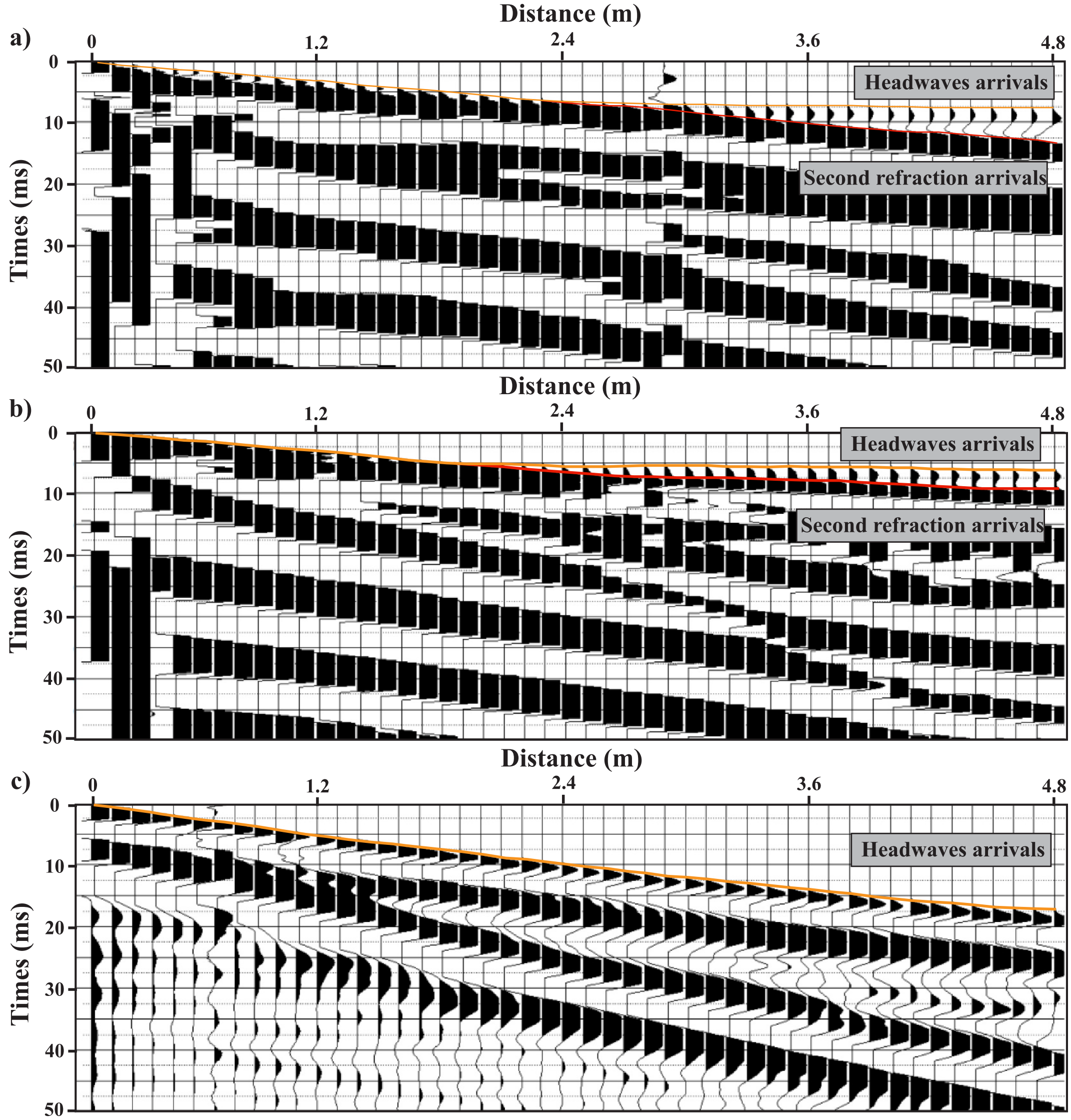

**Figure 7.** (a) A forward shot gather at the CSAD high-resolution 1 (HR-1) survey. (b) A forward shot gather at CSAD high-resolution 2 (HR-2) survey. (c) A forward point from the SJUW high-resolution (HR) survey. The red lines highlight first arrivals, while the orange lines indicate head wave arrivals. The shingled (head waves) and first arrivals from high-resolution surveys require a large gain to display them clearly. Refer to Figure 2 for survey locations.

### 3.3. $V_P$ Modeling

We employ reciprocal, time-term, and tomography techniques to obtain subsurface velocities and geometries. The techniques are chosen based on the investigation scale and data density. For the conventional seismic, we used the reciprocal time method, where more data density is required as this method requires multiple shots along the seismic profile (Figure 10). This method leverages the principle of reciprocity in travel times between seismic sources and receivers and allows more accurate velocity estimations and better subsurface layer geometry delineation (Palmer, 1981). Moreover, we used a seismic refraction tomography approach to obtain $V_P$ as a gradient per cell (tomogram) (Figure 11). The data density controls the tomogram of both the lateral and vertical $V_P$ resolutions. This technique utilizes picked arrival times as input without requiring preassigned layers and the direction of subsurface $V_P$ gradients (Azwin et al., 2013; Liu & Gu, 2012), making it more adequate for our study area. The picked arrivals underwent a reciprocity test before the inversion to avoid anomalous errors. Then, the obtained velocity model from the reciprocal method is employed as a simplified horizontal initial model to minimize the misfit of the $V_P$ inversion and define layer boundaries. The $V_P$ model is computed using the travel time plot, simulating the raypath of seismic waves through ray tracing (White, 1989). Velocities in each cell within the grid were adjusted iteratively to the lowest root mean square (RMS) error, minimizing the misfit between modeled and observed travel times (Figure 11). Meanwhile, at the high-resolution scale, we used the time-term approach, which is based on linear least squares and delay time analysis method to construct the $V_P$ section model (Boschetti et al., 1996; Palmer, 1981). This approach allows us to optimally determine the velocities of distinct subsurface layers by inverting seismic wave travel times using the picked arrivals (Figure 12). The time-term technique is a simple travel time analysis method and is considered a reliable technique where few shots are acquired (Iwasaki, 2002). The modeling was repeated three times using repicked data to compare the models against the errors from the manual picking, where minor changes were observed in the calculated velocities of less than 4%. Furthermore, the RMS error in all models had less than 1%–2% errors, validating that the model is very robust.

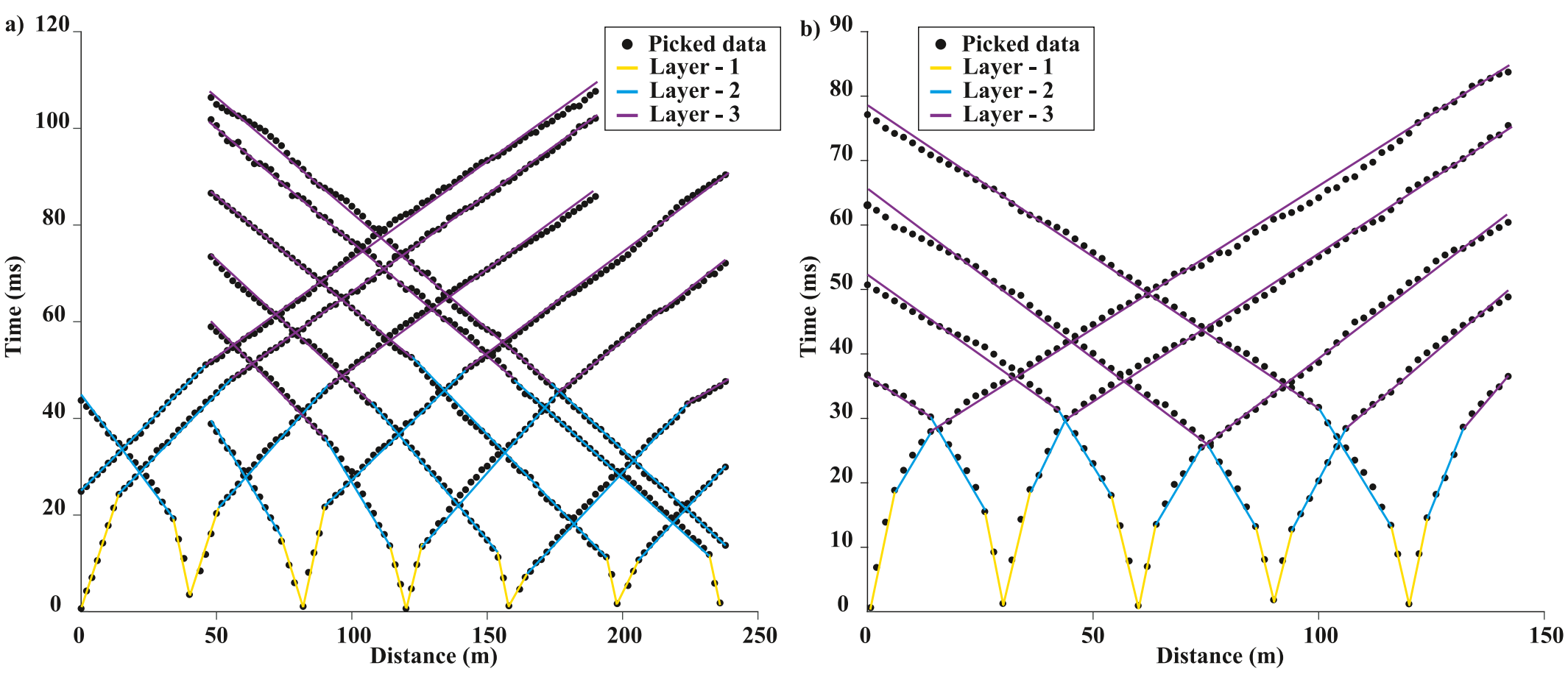

**Figure 8.** Selective time–distance curves constructed from the first arrival picking of the conventional seismic surveys. (a) CSAD conventional seismic survey. (b) SJUW conventional seismic survey. The color lines indicate the manually assigned layers based on best-fit regression slope changes.

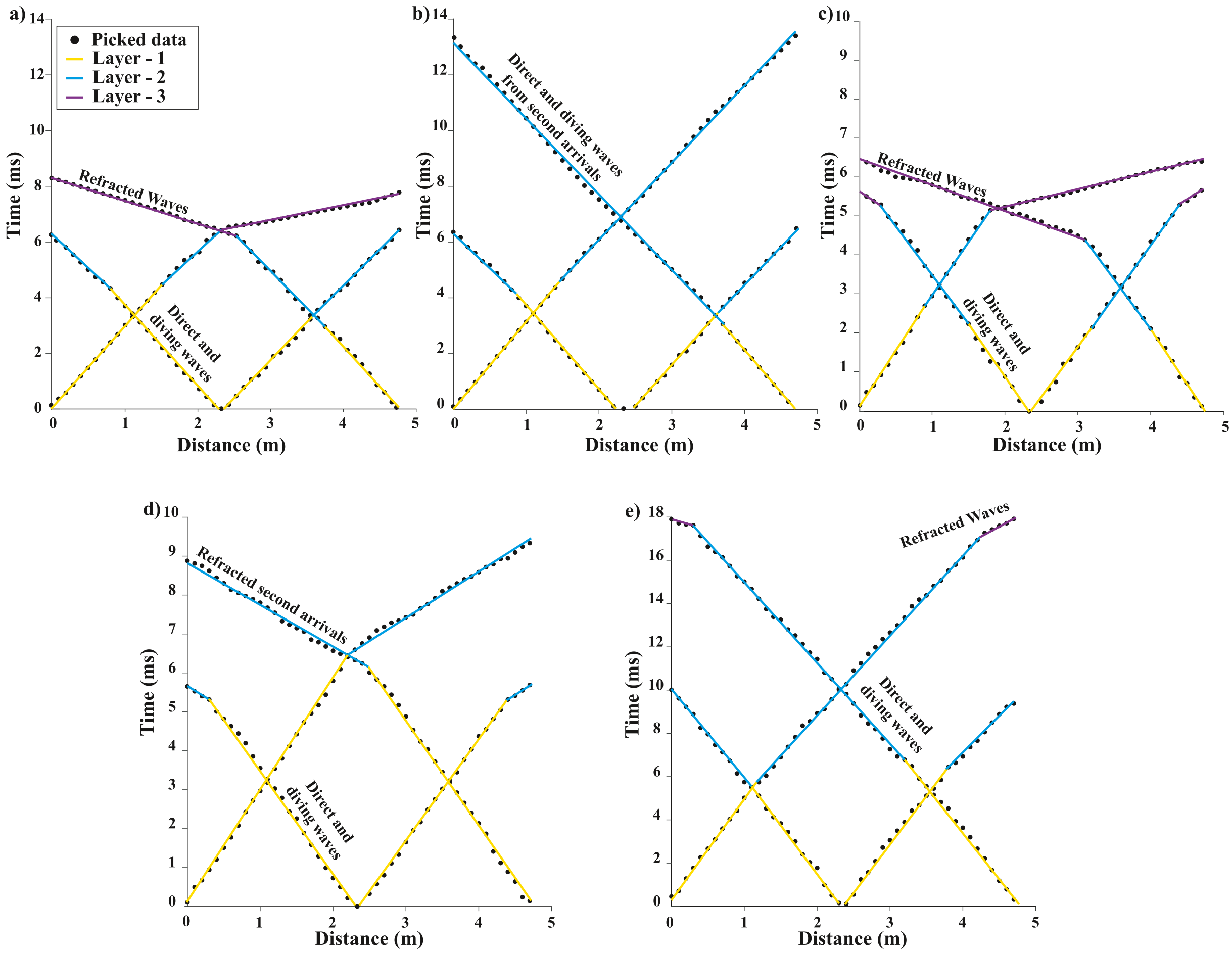

**Figure 9.** Time-distance curves obtained from high-resolution surveys utilizing forward, middle, and reverse shots. (a) Head waves arrivals picked from the CSAD HR-1 survey, (b) direct and second refraction arrivals picked from the CSAD HR-1 survey, (c) head waves arrivals picked from the CSAD HR-2 survey, (d) direct and second refraction arrivals from the CSAD HR-2 survey, (e) first arrivals picking from the SJUW HR survey. The color lines indicate the assigned layers based on best-fit regression slope changes.

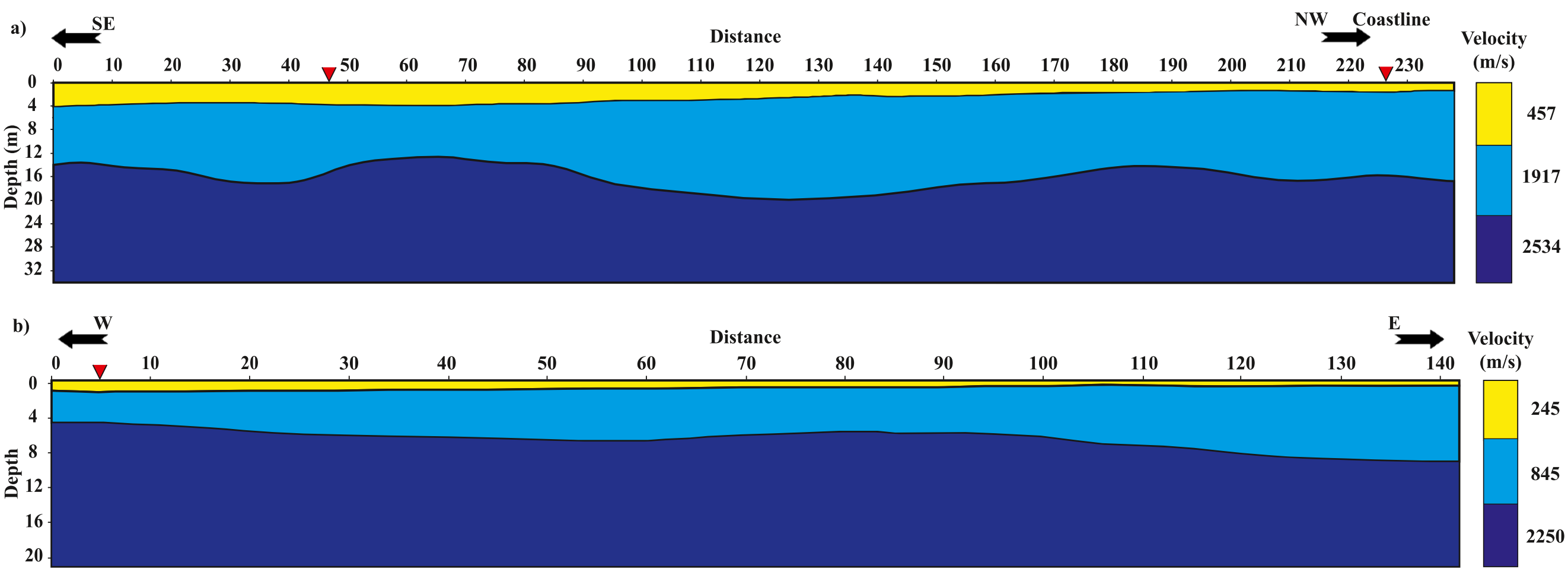

**Figure 10.** Average $V_P$ layer model obtained from the general reciprocal method of the conventional seismic data. (a) CSAD, (b) SJUW. Red triangles denote the locations of the high-resolution seismic experiments.

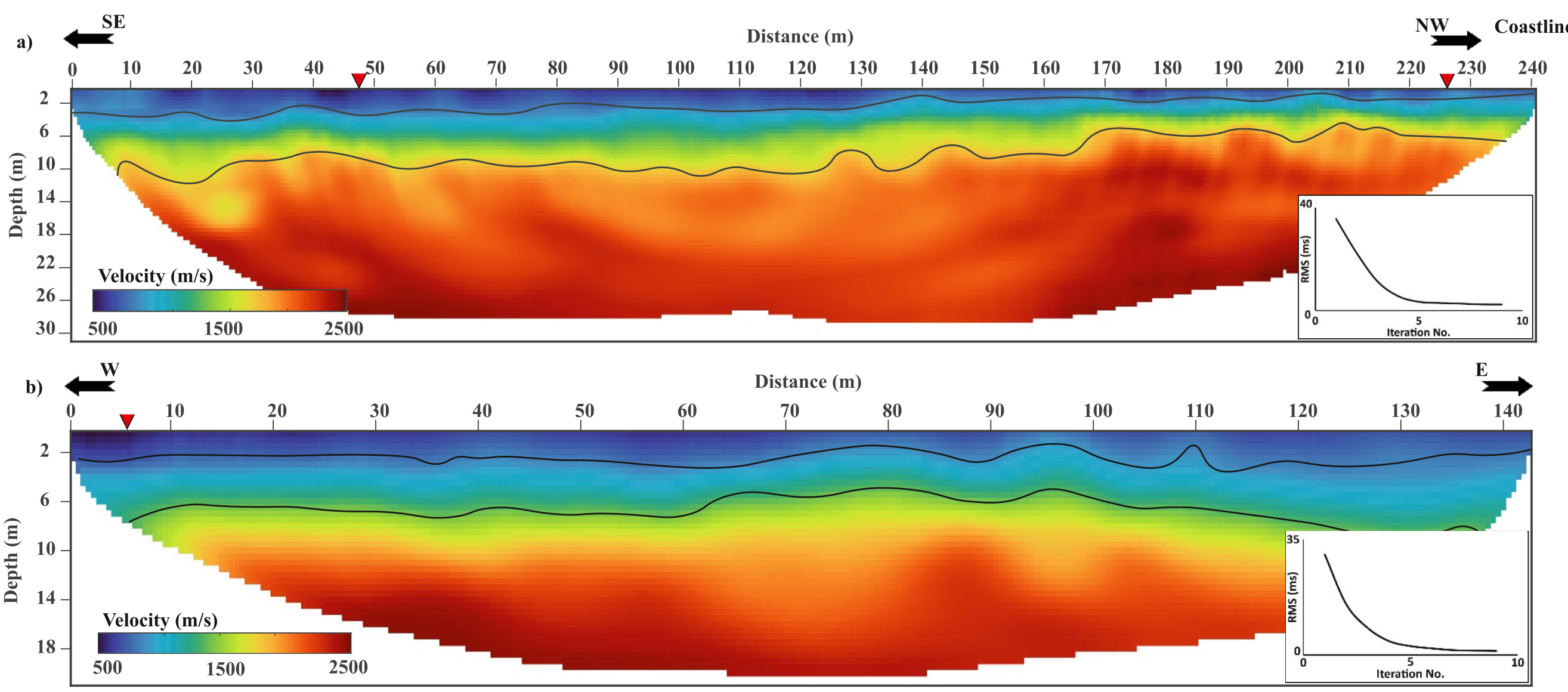

**Figure 11.** P-wave velocity inverted tomograms derived from conventional seismic surveys. (a) CSAD conventional seismic tomogram, and (b) SJUW conventional seismic tomogram. The bottom right figures show root mean square changes with the iteration numbers. Red triangles denote the locations of the high-resolution seismic experiments.

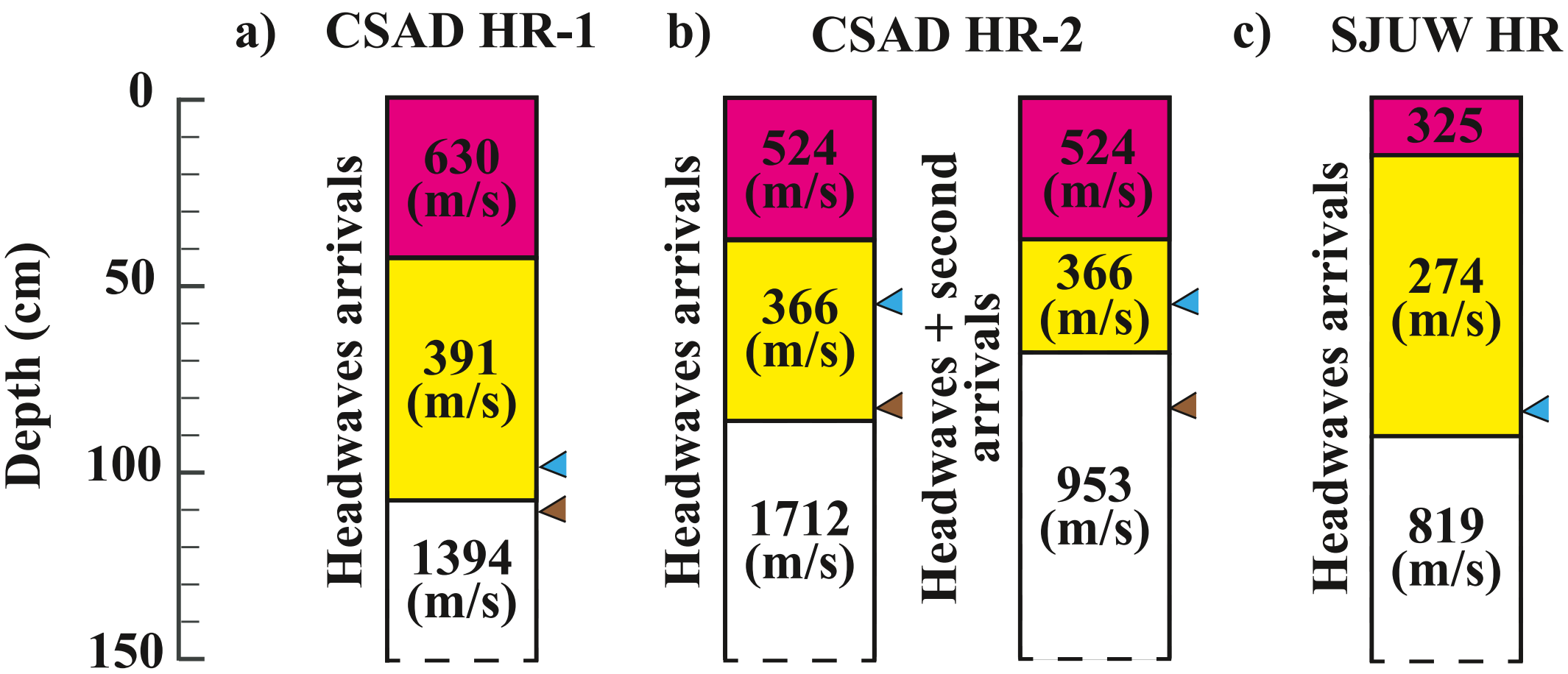

**Figure 12.** Schematic summary of the layered velocity models obtained from the high-resolution surveys of both sabkhas. The two models at CSAD HR-2 are obtained from head waves and second refraction of late arrivals picked arrivals in Figure 9. Water table from borehole observations are indicated by blue arrows, and hardground levels are shown by the brown arrows.

### 3.4. $V_P$ Estimation From Rock Physics Models

Rock physics modeling was used to estimate the $V_P$ in dry, partially, and fully saturated sabkha sediments and, hence, interpret the velocities obtained from seismic modeling (Figures 10–12). In the absence of shear velocity ($V_S$), Mavko et al. (1995) developed an approximate fluid substitution transform of $V_P$, operating directly on the P-wave modulus, $M$:

$$M = \rho V_P^2 = K + \frac{4}{3}\mu, \tag{1}$$

where $\rho$ is the bulk density, $K$ and $\mu$ are the bulk and shear modulus, respectively. The authors showed that the $M$ of a saturated porous media ($M_{\text{sat}}$) can be related to the modulus in the dry case ($M_{\text{dry}}$) using the following equation (Mavko et al., 1995):

$$\frac{M_{\text{sat}}}{M_0 - M_{\text{sat}}} \approx \frac{M_{\text{dry}}}{M_0 - M_{\text{sat}}} + \frac{M_f}{\phi\left(M_0 - M_f\right)}, \tag{2}$$

where $M_f$, $M_0$, $\phi$, $M_f$ are the fluid modulus, mineral modulus, porosity, and $M_f$ is the bulk modulus of the pore fluid, respectively. To estimate the $M_f$, we used the Batzle and Wang equations (Batzle & Wang, 1992) to determine the $K$ and $\rho$ of the water. We collected salinity and temperature measurements to calculate the $M_f$ and assumed an approximate hydrostatic pressure of 0.10 MPa. The estimated $K$ and $\rho$ of the water, among other data, are listed in Table 2. It is important to note that while the estimated brine bulk modulus (e.g., 3.48 GPa for CSAD) appears high relative to seawater (~2.3 GPa at 35 ppt salinity), it is appropriate for the hypersaline conditions encountered (241 ppt at CSAD). This is consistent with Batzle and Wang (1992) trends, which shows brine bulk moduli exceeding 3.5 GPa at salinities above 200 ppt and temperatures of 20–40°C. A similar calculation for SJUW (130 ppt, 28°C) yields $\rho = 1.089$ g/cm$^3$ and $K \approx 2.9$ GPa, matching the values listed in Table 2. The porosities of the sands below the sabkha sediments were obtained using a laboratory volumetric approach.

**Table 2**
*Input Parameters and Measurements Used in Equation 2*

| Parameter | CSAD | SJUW | Note |
|---|---|---|---|
| Temperature (°C) | 35 | 28 | Measured in situ conditions |
| Salinity (ppt) | 241 | 130 | Measured in situ conditions |
| Water density (kg/m$^3$) | 1,173 | 1,089 | Estimated based on Batzle and Wang (1992) |
| Water bulk modulus (GPa) | 3.48 | 2.90 | Estimated based on Batzle and Wang (1992) |
| $M_f$ (GPa) | 3.48 | 2.90 | $M_f$ is equal to bulk modulus for fluids based on Equation 1 since $\mu = 0$ |
| Porosity ($\phi$) | 0.38 | 0.32 | Measured in the lab for the upper unconsolidated sediments and assuming similar sand properties in deeper sections based on samples retrieved from study sites indicated in Figure 4 |
| $M_0$ (GPa) | 88 | 97 | A mixture of 24% anhydrite, 29% gypsum, 37% calcite and 10% quartz for CSAD and pure quartz mineralogy at SJUW and using elemental analysis for samples acquired from boreholes. The bulk and shear modulus for CSAD and SJUW are 53 and 26, and 37 and 44 GPa, respectively. In the context of the CSAD, we determine the average bulk and shear modulus of a minerals mixture using the Voigt-Reuss-Hill averaging method (Mavko et al., 2020) |

$M_{dry}$ was estimated using Equation 1, based on the $V_P$ values from the less evaporites affected zone (layer 2) and the dry bulk density ($\rho_{dry}$) derived from high-resolution models. It is assumed that the sands beneath the sabkha deposits exhibit comparable properties, as they were deposited in similar environmental conditions and the retrieved samples contain minimal evaporitic cement. The estimated $V_P$ is approximately 483 m/s for CSAD and 294 m/s for SJUW, measured just above the partially saturated layer. The dry bulk density was estimated using the following equation (assuming a grain density based on elemental composition):

$$\rho_{dry} = \rho_{grain}(1 - \phi) + \phi\, \rho_{air}, \tag{3}$$

$M_{dry}$ can then be calculated using Equation 1 before Equation 2 is used to estimate $M_{sat}$. Equation 3 was also used to determine the $\rho$ of saturated media, but water was used instead of air density. Finally, $M_{sat}$ and $\rho_{sat}$ were used to calculate the $V_P$ of the fully saturated sands below the sabkha sediments that occur at average depths of 3.5 and 5 m at CSAD and SJUW, respectively, which can be compared with the $V_P$ obtained from the conventional seismic to confirm that it is due to the full water saturation and not a lithological interface.

We used the Ruess average to calculate $M_{f\text{-fully-saturated}}$ and $M_{f\text{-partially-saturated}}$ for the multi-mineral composite and partial saturation. According to the Reuss model, the average is calculated based on the assumption that the stress is uniform across the different phases, but the strain can vary (Hill, 1963). For elastic moduli, the Reuss average is given by:

$$\frac{1}{M_f} = \frac{S_w}{M_{f\ brine}} + \frac{S_{air}}{M_{f\ air}}, \tag{4}$$

where $M_f$ is the effective bulk modulus, $S_w$ and $S_{air}$ are the saturation fractions of water and air, respectively, and $M_{f\ brine}$, and $M_{f\ air}$, are the bulk moduli of water and air. Both the water and air moduli, representing a two-phase fluid system within the pore space, are averaged according to their respective volume fractions in the matrix to obtain the effective bulk modulus of the fluid.

### 3.5. Static Correction

To calculate the static correction, the weathering layer is typically separated and substituted by another layer with a constant or slowly variable velocity known as the replacement velocity, often referred to as the datum velocity, elevation velocity, or sub-weathering (Cox, 1999). The conventional approach commonly used is replacing the weathered layer with a layer with an assumed $V_P$ of 300 m/s (Steeples et al., 1990). To compute the static correction from a constant replacement velocity, the replacement velocity is determined using the $V_P$ profile measured at the site. The inverse of the slope of the first arrival picks is used to calculate the $V_P$. The static correction $t_{SC}$ is calculated as:

$$t_{SC} = \sum_{i=1}^{i=x} \frac{2H}{V_R}, \tag{5}$$

where $i$ is the layer number, $H$ is the thickness of the sabkha sediments and $V_R$ is the replacement $V_P$ from the conventional seismic survey. Factor 2 is introduced to compute the source and receiver sides.

To compute the static correction using seismic tomography, we took $V_P$ values of the travel time tomogram, known as tomostatics (Zhou, 2014). The travel time tomogram produced from conventional seismic data (Figure 11) is distributed into square-grid cells of $0.5 \times 0.5$ m$^2$, assuming that the $V_P$ is constant within each grid cell. To determine the static correction, we assumed that the seismic energy propagates in a vertical direction, hence a 1-D $V_P$ model extracted from a midpoint of the tomogram. The travel time from the sediment surface to the datum (the source side) and back to the top surface (the receiver side) is then calculated using Equation 6:

$$t_{SC} = 2h \sum_{Z=0}^{Z_{max}} \frac{1}{V_Z}, \tag{6}$$

where $Z$ varies from $Z = 0$ m at the ground surface to $Z_{max}$ corresponding to the top of the fully saturated layer, $h$ is the thickness of each tomography cell (0.5 m in our case), and $V_z$ is $V_P$ at each cell.

## 4. Results

### 4.1. $V_P$ Modeling of Conventional Seismic Surveys

At the CSAD, three layers were defined from the conventional seismic data using the slope analysis of picked time-distance curves (Figure 8a) and computed velocity model from the reciprocal method (Figure 10a). The model exhibits an average velocity of 457 m/s for the first layer, 1,917 m/s for the second layer, and 2,534 m/s for the third layer. The upper layer is significantly thicker away from the coastline and gradually thins toward the coastline. The boundary between the first and second layers shows a smooth interface, which pinched out toward the end of the section, contrasting with the more intricate and complex geometry observed at the third layer. In the SJUW case, three layers were distinguished (Figure 10b); the topmost layer is notable for its minimal thickness, averaging 1.5 m with a $V_P$ of 245 m/s. An asymmetrical thickness distribution is evident in the second layer, with the greatest thickness extending toward the east, corresponding to a velocity $V_P$ of 845 m/s. The third layer becomes apparent at a depth of 6 m, characterized by a $V_P$ of 2,250 m/s. Furthermore, the tomograms from both sabkhas give a better insight into the velocity changes within the same layer in vertical and horizontal directions (Figure 11). In the CSAD tomogram (Figure 11a), the shallowest layer exhibits a velocity ranging between 450 and 780 m/s, where this layer is pinching out toward the coastline direction and has low velocity localized at certain points along the section and right below the surface. A more homogenous second layer is characterized by a velocity ranging between 780 and 1,750 m/s. Meanwhile, the deepest layer shows velocities ranging between 1,750 and 2,480 m/s, where this layer exhibits a high degree of heterogeneity with large changes in the lateral direction. Moreover, the CSAD's first and third layers exhibit significant vertical and lateral heterogeneity, indicating their high degree of complexity. The tomogram of SJUW shows less complexity in shallow parts when compared to CSAD (Figure 11b). The shallowest layer exhibits a velocity ranging between 380 and 880 m/s, where this layer exhibits a relatively similar lateral thickness and thinning in localized parts. A more homogenous second layer is characterized by a velocity ranging between 880 and 1,620 m/s. It exhibits a similar thickness with a slightly thicker part toward the end of the profile in the eastern direction. Meanwhile, the deepest layer shows velocities ranging between 1,620 and 2,465 m/s, where this layer exhibits a high degree of heterogeneity with large changes in the lateral direction. Additionally, this layer shows a localized lower velocity zone of a synclinal shape toward the deeper parts of this layer.

**Table 3**
*Summary of Obtained Velocities of Each Layer Through High-Resolution, Conventional Seismic, and Rock Physics Approaches, Indicating the Saturation Degree for Partially Saturated Layers*

| | CSAD $V_P$ (m/s) | | | SJUW $V_P$ (m/s) | | |
|---|---|---|---|---|---|---|
| Layer | High-resolution | Conventional seismic | Rock physics | High-resolution | Conventional seismic | Rock physics |
| Dry | 483 | 457 | 468 | 294 | 245 | 278 |
| Partially saturated | 953 | – | 907 ($S_w$ = 99.3%) | 819 | 840 | 840 ($S_w$ = 99.2%) |
| Fully saturated | – | 1,906 | 2,016 | – | 2,250 | 2,162 |

### 4.2. $V_P$ Modeling of High-Resolution Seismic Surveys

At the high-resolution scale, the time-term technique was used to construct average velocity models (Figure 12) using time-distance curves from head waves and second refracted arrivals at CSAD (Figures 12a and 12b), and head wave arrivals only from SJUW (Figure 12c). We obtained high-resolution velocity models to resolve the complexity of the upper few tens of centimeters associated with evaporitic facies, hardground, and shallow water table, to obtain the average velocity of dry (weathered) and wet layers. Additionally, we computed the velocities using the picked time-distance curves from the second refraction arrivals at HR-2 (Figure 12b). In the case of CSAD, the models obtained from high-resolution survey head wave arrivals picking identified three layers (Figures 12a and 12b). The topmost layer is characterized by an average velocity of 577 m/s. Below it, a second layer with an average velocity of 379 m/s and a third layer with an average velocity of 1,553 m/s.

At HR-1 location, the second refracted arrivals wave picking approach did not recognize any additional layers at site HR-1 (Figure 9b). However, an additional distinct layer was recognized at HR-2 using second refracted waves (Figure 9d). The first and second layers are characterized by average velocities of 577 and 379 m/s, respectively, while the third layer exhibits a higher velocity of 1,553 m/s from head wave arrivals and 953 m/s from second refracted arrivals (Figures 12a and 12b). In the head wave arrivals model, the deepest layer demonstrates significantly greater velocities than the second late arrivals. Furthermore, at SJUW HR-1, three layers were distinguished with velocities of 325, 274, and 819 m/s, respectively, where the uppermost layer is very thin, almost 5 m (Figure 12c).

### 4.3. $V_P$ Estimation Using Rock Physics Models

The computed velocities (Table 3) from rock physics match those obtained from seismic refraction surveys. The dry layer exhibits significantly lower velocities compared to the partially and fully saturated layers, reflecting the lack of fluid content and higher porosity. Partially saturated layers show intermediate velocities, where the presence of fluid increases the wave speed but does not reach the high values seen in fully saturated conditions. For the fully saturated layer, the seismic velocities exceed 1,500 m/s, indicating the presence of water filling the pore spaces completely, which enhances wave propagation. The minor discrepancies between rock physics and seismic refraction velocities in the partially and fully saturated zones can be attributed to fluid distribution variations and the subsurface materials' heterogeneous nature. These results indicate that the velocities observed in the seismic refraction models are due to a partially saturated zone occurring in both sabkhas and are a unique feature in this environment. The fully saturated layer (>1,500 m/s) is only recognized in conventional seismic as they are found at deeper depths than the investigation scale of the high-resolution survey. Moreover, the rock physics modeling results support the partial saturation zone in the sabkha sediments, eliminating the possibility of being a lithological boundary. It is evident that a water saturation ($S_w$) of approximately 99.2%–99.3% yields a velocity similar to observed values from the seismic refraction models. We assessed the variations in the final computed $V_P$ by accounting for uncertainties in core-derived input parameters, such as porosity (±2%), mineralogical composition, and water salinity. This led to relative errors in $V_P$ values of approximately 18–32 m/s in dry layers and up to 121–168 m/s in fully saturated layers, yielding an overall variation of around 7% or less compared to the obtained velocities. The relative effect remains within acceptable limits for both dry and saturated conditions.

**Figure 13.** Scanning Electron Microscope images at different resolutions of the anhydrite facies obtained from CSAD BH-1 at a depth of 23 cm. The degree of evaporitic maturity indicated by the dominating anhydrite (AN) crystals completely replaced most of the original sediments and created a dense matrix of elongated anhydrite (chicken wireline anhydrite) crystals with very low porosity. Additionally, some minor gypsum cement (CE) exists within the sample.

## 5. Discussion

### 5.1. Impact of Lithological Variations

Both conventional and high-resolution data from the CSAD exhibit shingled wave arrivals, leading to amplitude variations with offset (Figures 6a, 7a, and 7b). This shingling phenomenon is caused by the presence of a very high-velocity thin refractor (Cassinis & Borgonovi, 2006; Donato, 1965; Press et al., 1954). Shingling manifests as the gradual disappearance of one or more refracted waves into lengthy offsets, rendering it challenging to track the first arrivals. In practice, seismic data often feature a combination of shingling and non-shingling records, with a segment of shingling following or preceding a region of second arrivals, reflecting the continuous or discontinuous nature of geological formations (Diggins, 2016; Sun et al., 2018). In the CSAD case, the shingling of first arrivals results from the presence of a thin hardground (<50 cm) layer beneath the sabkha sediments at a very shallow depth (<1 m). Shingled seismic arrivals are undesirable for seismic refraction studies due to their thin-layer nature and noncontinuity, as indicated by their significantly higher velocity (Cassinis & Borgonovi, 2006). Consequently, for conventional seismic data, we avoid shingled waves when picking the first arrivals and instead focus on the later second-refracted wave arrivals, which represent the continuous arrivals of a thick, gradually increasing velocity layer. Furthermore, the tomogram models (Figure 11) show high levels of heterogeneity in the CSAD subsurface compared to SJUW. CSAD is characterized by a more complex stratigraphic and geological history caused by sea level fluctuations during the past 10 kyr (Kirkham & Evans, 2019).

The high-resolution method effectively resolved the complexity of the sabkha environment, clarified the influence of evaporitic deposits, and delineated the main stratigraphic zones, including the presence of hardground, based on their seismic properties. A near-surface high-velocity layer identified in the high-resolution models (Figure 12) exhibits significantly higher velocities in CSAD compared to SJUW within the upper few tens of centimeters, ranging from 524 to 630 m/s in CSAD and approximately 325 m/s in SJUW. This high-velocity layer is a result of greater evaporitic facies growth where more salt cement is abundant between the grains (Figure 13). Additionally, this zone is characterized by a thin surface of halite, followed by an anhydrite layer and gypsum (Figure 4). The high-velocity layer is followed by a lower-velocity layer at 38 and 36 cm depth at HR1 and HR2, which correlates to a less dominant salt zone, mainly composed of sands and smaller portions of evaporitic minerals, where average velocity ranges between 274 and 391 m/s.

The third layer in the high-resolution models computed from the head waves arrival time-distance curves (Figures 9a and 9c) exhibited average velocities ranging between 1,394 and 1,712 m/s (Figure 12). The head waves arrivals for the third layer in the high-resolution models match the hardground depths from borehole observations of 103 and 82 cm at HR-1 and HR-2, respectively (Figure 12). The hardground layer displays velocity variations as a result of diagenetic features since the last sea level rise, where the hardground is shallower, thicker, and more lithified toward the sea, as reported where it outcrops in the lagoon bottom (Paul & Lokier, 2017). The hardground is composed of bioclastic fragments, quartz, and calcite grains with a mixture of aragonite and calcite extensive cement (Figure 14), enclosing most of the intragranular porosities. This explains the high $V_P$ values obtained from the seismic models and validates the interpretation of the data. These results correlate with the observations of lithological successions obtained from both CSAD HR-1 and HR-2 and SJUW boreholes.

**Figure 14.** Scanning Electron Microscope images at different resolutions of hardground fragments obtained from CSAD BH-2 at a depth of 85 cm. The sample is composed of excessive calcite cement (CE) filling the intergranular pores, large bioclastic fragments (BI), and carbonate mud (MU). The dense cementation shows a high degree of lithification of the rock.

The high heterogeneity of sabkha, coupled with the presence of evaporitic facies and the nature of the thin hardground layer, leads to significant attenuation. This results in the rapid decay of shingling waves, posing challenges for their detection using conventional seismic surveys. The high-resolution surveys had a wavelength of investigation sufficient to resolve the hardground and found that resolving CSAD hardground through conventional seismic data is impossible due to relatively larger wavelengths and, therefore, lower detection and resolution. This explains the disappearance of the shingled waves after five geophones (10 m), as most of the high frequencies are attenuated. Thus, resolving hardgrounds is impossible in conventional seismic surveys and typically results in errors in seismic velocity studies.

The values obtained from rock physics and seismic modeling validate and explain the velocity contrasts and changes, where a sufficient concentration of evaporitic minerals close to the surface can cause a higher velocity (>300 m/s) than typical soils or weathered layers within the upper first meter. Additionally, in CSAD, the presence of hardground results in a high-velocity layer (>1,500 m/s), but due to the thin nature of the layer, it results in shingled arrivals as layer velocity is out of trend and overlain and underlain by lower velocity layers. The sabkha velocities tend to deviate from typical sandy sediments, where $V_P$ is generally increase with depth due to the progressive rise in overburden stress, as discussed by Bachrach et al. (2000), , and demonstrated in Al-Shuhail et al. (2018). Moreover, conventional seismic methods are inadequate for capturing the intricacies of the uppermost one to two m (Ackermann et al., 1986; Bery, 2013). This zone is defined by a thin and heterogeneously distributed bed of evaporites exhibiting horizontal and vertical variations. The CSAD displays a greater degree of maturation in evaporitic facies, recording various evaporitic facies. In contrast, the SJUW features a more limited presence of evaporites and less maturity (Figure 4).

### 5.2. Partially and Fully Saturated Zones

The conventional seismic velocity models for CSAD and SJUW identified a fully saturated layer, characterized by a $V_P$ velocity near to 1,500 m/s at depths greater than the expected 3–5 m range (Figure 10). The $V_P$ of shallow subsurface saturated layers may be lower than 1,500 m/s in some regions, such as the tropics, owing to swiftly increasing water levels caused by severe rains, resulting in air entrapment (Desper et al., 2015). Other possibilities for lower-than-expected $V_P$ values may be shallow water tables in unconsolidated sedimentary strata (Pasquet et al., 2015). In some areas, shallow groundwater aquifers exhibit low velocities (800–1,400 m/s) (Abdelrahman et al., 2017; Azhar et al., 2019). Moreover, the velocity found in the high-resolution models at depths matching the water table from borehole observations was found to be less than the velocity of water (1,500 m/s), characterized by an average velocity of 953 m/s at CSAD and 835 m/s at SJUW (Figures 12b and 12c). Additionally, a comparable layer exhibiting an average velocity of 845 m/s was identified by the SJUW conventional seismic survey at depths consistent with those observed in the high-resolution experiment at the same location. Bachrach and Nur (1998) and Bachrach et al. (1998) observed a similar feature during their investigation of beach sands to determine the water table depth using high-resolution seismic refraction and reflection. These authors observed that saturation significantly affects seismic velocity, as the first-layer velocities from both refraction and reflection surveys decreased during high tide and gradually increased as the sand drained. Both wave types were influenced by partial saturation, exhibiting sensitivity to fluid flow history. Their findings were further validated through a rock physics analysis. Furthermore, the inland sabkha of Jayb Uwayyid (SJUW) in eastern Saudi Arabia was previously investigated by Al-Shuhail and Al-Shaibani (2013) using five seismic refraction profiles with a geophone spacing of 5 m. They identified three layers where the shallowest layer has an average $V_P$ of 700 m/s and thickness of 12 m and was interpreted as a partially saturated layer. The lower layer had an average $V_P$ of 2,300 m/s, interpreted as a fully saturated partially lithified rock, while the deepest layer had a $V_P$ of 3,350 m/s and was interpreted as bedrock. They described this layer in their model as a partially saturated layer but failed to distinguish the boundaries between dry and partially saturated layers, which is attributed to the use of widely spaced geophones. Therefore, this layer is a partially saturated layer overlain by the dry sabkha sediments. At CSAD, the detection of a partially saturated layer has failed using conventional seismic methods (Figure 10a), mostly due to the very thin transition zone between the partially saturated and fully saturated zones. Furthermore, the first high-resolution experiment at CSAD did not detect any partially saturated layer, consistent with the conventional seismic results, where the fully saturated layer interface appears at greater depth relative to location HR-2. This suggests that the partial saturation interface may lie deeper than at location HR-1 (Figure 10a). In both cases, high-resolution experiments detected the partially saturated layer (>850 m/s). Meanwhile, conventional seismic has succeeded in mapping the partially saturated layer only in SJUW but has failed to do so in CSAD. The precise matching of rock physics-derived velocities with those from high-resolution and conventional seismic surveys reinforces the reliability of these methods in characterizing subsurface conditions, particularly in complex environments like sabkhas where saturation levels vary.

Figure 15. Schematic of the typical saturation trend in sabkha against depth. Based on the seismic refraction results, three main zones are recognized: moist (vadose) zone as a result of capillary fringe and evaporation directly from the water table, partially saturated (phreatic) zone directly linked to the water table level, which has some trapped gas in the pore space, and the fully saturation zone.

The velocity models derived from high-resolution seismic data (Figure 12) revealed a partially saturated layer at depth, consistent with borehole observations inferred from later seismic arrivals. Figure 15 explains the transition between moist, partially, and fully saturated zones in the sabkha subsurface. As illustrated, the shingled arrivals observed in the conventional seismic survey are attributed to the presence of hardground, which is more effectively detected and characterized using high-resolution seismic techniques. Meanwhile, the refracted second arrivals are a result of the partial saturation zone occurring above and below the hardground and characterized by a lower velocity compared to the hardground. Despite observing the water table at a depth of less than 1 m, the measured velocity suggests the presence of some air/gas, so the porous media may not be considered 100% saturated. Solazzi et al. (2021) investigated the effect of capillary forces on the partially saturated layers using rock physics modeling, additionally, Azhar et al. (2019) documented low-velocity saturated sands at the water table level and explained this by the changes in the water table level affected by rains and low permeability lithologies causing entrapment of gas and potentially causing the water table to appear deeper when using seismic refraction method. The entrapment of gas in the pore space may be attributed to several factors, including the rise of the paleowater table during sabkha formation with the presence of impermeable zones leading to the creation of gas pockets (Al-Farraj, 2005; McKay et al., 2016); diagenetic processes, including sulfate reduction and algal mate decomposition or microbial methanogenesis (Wood et al., 2002); or the combination of both processes. It is worth noting that during the drilling of the borehole, a significant observation was the emergence of bubbles, as the borehole was filled with water up to the water table level. The rate of water ascent varied between locations due to permeability changes in the underlying sediments. The gradual increase in the pore-space saturation does not generate refracted head waves in the absence of a sharp velocity interface (Al-Shuhail & Al-Shaibani, 2013; Sjogren, 2013). Therefore, we expect the third layer to be a lithological interface attributed to more consolidated, fully saturated sand. The deeper sandy layers below the sabkha sediments in the conventional surveys have an unconsolidated nature, yet the recorded $V_P$ velocities exceed 1,900 m/s, mostly attributed to the salinity content fluctuation of the groundwater as observed in the velocity tomograms (Figure 11) and previously documented by Van Dam et al. (2009). The vertically structured anomalies observed in the third layer in SJUW may be attributed to weaknesses in the elastic modulus of the rock due to diagenetic features such as karst or cavities, as those layers are of a carbonate nature. Therefore, sabkha sediments are highly heterogeneous, with strong variations in thickness and salt concentrations.

**Table 4**
*Summarized Static Correction Models and Their Corresponding Errors for High-Resolution Seismic Correction*

| Static correction type | CSAD (ms) | SJUW (ms) |
|---|---|---|
| Conventional replacement | 26.7 | 44.7 |
| Constant replacement | 17.7 | 23.4 |
| Tomostatic | 13.9 | 18.7 |
| Combined conventional and high-resolution | 10.7 | 23.1 |

### 5.3. Static Correction Implications

We compared the velocities obtained from static correction approaches to the high-resolution surveys, assuming the top of the fully saturated layer is the first reflector and the static correction of the low-$V_P$ sabkha sediments and the partially saturated layer need to be determined (Al-Shuhail, 2002; Hanafy et al., 2020). In conventional seismic data, the computed average velocities of dry and partially saturated sediments are used as replacement velocities. The upper few meters of the conventional CSAD seismic belt had a $V_P$ of 452 m/s and a thickness of 4 m, while SJUW had two layers with $V_P$ values of 245 and 845 m/s with thicknesses of 1.3 and 5.4 m, respectively (Figure 10). The computed static correction using the replacement average velocity model at a mid-point corresponding to the high-resolution survey and using Equation 5 is 17.7 ms for CSAD and 23.39 ms for SJUW. Moreover, the $V_P$ values in the tomograms between the ground surface and the top of the fully saturated layer fall between 440 and 790 m/s for CSAD and between 320 and 840 m/s for SJUW (Figure 11), where each cell in the tomogram is 0.5 m in length and Equation 6 has been utilized to obtain the static correction. The calculated static corrections using the tomostatic approach are 13.88 ms in the case of CSAD and 18.74 ms for SJUW. Furthermore, we incorporated both the values from the high-resolution survey and the conventional layered seismic model results. We used Equation 5 to calculate the travel time from the ground surface to the top of the fully saturated sands underlying the sabkha sediments. Where $V_z$ is the average $V_P$ (Figures 12 and 14) for each layer, and the depth $z$ varies based on the thickness of each layer. The calculated static correction from the high-resolution and conventional seismic data was 10.7 ms for CSAD and 23.1 ms for SJUW. If we take the high-resolution result as the reference, as it can resolve the very small-scale details that were not achieved in conventional seismic, and compare it to other static corrections approaches, the error differences are as follows in Table 4.

The low velocities observed at extremely shallow depths, which are often ignored, can cause substantial inaccuracies in static corrections and, as a result, large shifts in the placements of deeper reflectors (Cox, 1999; Yilmaz, 2016; Zhou, 2014). The differences in the static correction values are due to the differences in seismic resolution and sensitivity to the boundary contacts. Increased heterogeneity within the geological subsurface corresponds to a heightened level of complexity, necessitating a more detailed resolution approach to unravel these intricate features effectively. High-resolution techniques excel in this regard, providing superior delineation.

Due to the high heterogeneity of sabkhas, static corrections can be substantially improved using high-resolution data, which delineates subtle variations more accurately in layer boundaries and enhances correction precision. In conventional seismic data, it is generally assumed that there is no partially saturated zone, and the first reflector is assumed to be shallow, leading to an increase in the error of the applied corrections.

In the CSAD case, the replacement velocity from conventional seismic data showed a very high error when compared to that from high-resolution data, mostly as a result of failure to detect the hardground or the partially saturated layer. Meanwhile, at the SJUW the replacement velocity had a small error as it succeeded in mapping the partially saturated layer, but the error was mostly a result of inaccurate depth mapping. The tomostatic approach had relatively similar errors for both cases.

Our results show that static correction is very sensitive to changes in $V_P$ variations within the uppermost zone, which can drastically affect the correction and imaging of the deeper subsurface. Additionally, conventional seismic methods may be limited in resolving complex near-surface geological conditions, which impact static correction estimates due to variations in sabkha lithologies, such as evaporitic mineral content, hardground presence, and differing degrees of sediment saturation. Static correction utilizing both high-resolution and conventional seismic data has successfully addressed the velocity changes within the uppermost parts of the sabkha. This highlights the complexity associated with velocity changes within the first few meters of sabkha sediments. It is necessary and recommended that high-resolution seismic data be acquired at CSAD when carrying out a seismic survey of mature developed sabkha. However, conventional seismic methods can be acceptable for immature sabkha, such as SJUW, because of their low degree of heterogeneity, which is less challenging than that of mature sabkha. Therefore, the complexity of the uppermost layers is critical, as it substantially impacts the accuracy of static corrections.

The current study emphasizes the necessity of obtaining high-resolution seismic data in heterogeneous mature coastal sabkha environments to detect velocity changes in the uppermost layers precisely. However, uncertainties persist in seismic survey results, which could be addressed by drilling a well into both sabkhas to explore their lithological sequence. Furthermore, conducting borehole seismic surveys can refine velocity measurements at higher resolutions and under in situ conditions, while obtaining core samples would allow for a detailed investigation of the physical properties of underlying sediments sequences. These suggested avenues for further research would significantly enhance our understanding of sabkha formations and their implications for seismic velocity applications.

## 6. Conclusions

A seismic investigation study was carried out at the mature CSAD and immature SJUW, using conventional seismic profiles with geophones spaced 2 m apart, and incorporated a high-resolution survey using geophones that were closely spaced (10 cm) to map the velocity at two different scales. The main findings of the study are:

- Shallow boreholes obtained from both sabkhas delineated the salt concentrations in the upper few centimeters, and the water table depth was less than 1 m. However, the computed velocity models displayed velocities less than the velocity of water (<1,500 ms) or showed a saturated layer at deeper depths than did the borehole observations. Those layers were verified as partially saturated zones, which is a unique feature of the sabkha environment due to the entrapment of air as a result of different mechanisms, leading to lower velocities than typically expected for saturated sediments.
- The presence of evaporitic facies in the shallowest layer leads to a higher velocity than in typically weathered layers. In the case of mature CSAD, the velocity ranged between 524 and 630 m/s. Meanwhile, the developing SJUW exhibited a velocity of 325 m/s. Therefore, the velocity increase varies with the degree of sabkha maturity. In the immature SJUW, the velocity of the sabkha sediments is close to the typical weathered layer velocity (≈300 m/s). Meanwhile, in mature CSAD, the velocity of the first layer exceeds 500 m/s. Hence, replacement velocities are not applicable in mature sabkhas.
- The presence of a thin lithified hardground at the CSAD shallow surface leads to shingled arrivals in the conventional seismic survey, which are impossible to detect using traditional seismic methods. The hardground exhibited a velocity ranging between 1,394 and 1,712 m/s due to a high degree of cementation in the pore space.
- We applied conventional, replacement, and tomogram approaches for static correction and compared them to high-resolution corrections at mature coastal and immature inland sabkhas. A relatively small change was found in the conventional static correction methods compared to the high-resolution approach in the immature inland sabkha, as the small error is mostly due to computational error and is not resolution- or method-related. However, in the mature coastal sabkha, the difference between high-resolution is greater, where the large variations in the near-surface complexity were averaged in the conventional approach. Therefore, high-resolution seismic is necessary for mature sabkhas, compared to immature sabkhas, where replacement velocity is applicable.

## Data Availability Statement

The CSAD and SJUW seismic data sets were collected by Khalifa University and King Fahd University of Petroleum and Minerals (KFUPM), respectively, and can be obtained from this depository: Eleslambouly (2024). SeisImager/2D, a trademark of Geometrics (Software—seismic analysis was made with SeisImager/2D version 4.0, available in https://www.geometrics.com/software/seisimager-2d/). The seismic refraction tomography Matlab code (DSISoft) (Beaty et al., 2002).

**Acknowledgments**
The authors would like to acknowledge both Khalifa University and King Fahd University for supporting the acquisition of the data in this work.